# Carrier Trapping by Oxygen Impurities in Molybdenum Diselenide

*Ke Chen[1], Anupam Roy[2], Amritesh Rai[2], Amithraj Valsaraj[2], Xianghai Meng[1], Feng He[1,4], Xiaochuan Xu[3], Leonard F. Register[2], Sanjay Banerjee[2], Yaguo Wang[1,4]\**

1. Department of Mechanical Engineering, The University of Texas at Austin, Austin, TX 78712, USA
2. Microelectronics Research Center and Department of Electrical and Computer Engineering, The University of Texas at Austin, Austin, TX 78758, USA
3. Omega Optics, Inc., Austin, TX 78757, USA
4. Texas Materials Institute, The University of Texas at Austin, Austin, TX 78712, USA

*Corresponding Author: yaguo.wang@austin.utexas.edu

Abstract

Understanding defect effect on carrier dynamics is essential for both fundamental physics and potential applications of transition metal dichalcogenides. Here, the phenomenon of oxygen impurities trapping photo-excited carriers has been studied with ultrafast pump-probe spectroscopy. Oxygen impurities are intentionally created in exfoliated multilayer $MoSe_2$ with $Ar^+$ plasma irradiation and air exposure. After plasma treatment, the signal of transient absorption first increases and then decreases, which is a signature of defect capturing carriers. With larger density of oxygen defects, the trapping effect becomes more prominent. The trapping defect densities are estimated from the transient absorption signal, and its increasing trend in the longer-irradiated sample agrees with the results from X-ray photoelectron spectroscopy. First principle calculations with density functional theory reveal that oxygen atoms occupying Mo vacancies create mid-gap defect states, which are responsible for the carrier trapping. Our findings shed light on the important role of oxygen defects as carrier trappers in transition metal dichalcogenides, and facilitates defect engineering in relevant material and device applications.

In the past few years, layered materials transition metal dichalcogenides (TMDs) have attracted numerous research interests due to their extraordinary properties, such as direct bandgap in monolayer samples[1], stable excitons and trions at room-temperature[2], superior immunity to short channel effects[3], and strong spin-valley coupling[4], which make TMDs promising for applications in nano- and flexible electronics, photonics, and valleytronics. For many of these TMD-based devices, it is essential to understand the physics of carrier-related interactions[5], and to acquire carrier transport properties that determine the key performance metrics (bandwidth, responsivity), such as carrier lifetime[6], diffusion coefficient[7] and mobility[8].

Defects inside TMDs or from their environments (substrates, surface adsorbates) can interact with excitons/carriers, and significantly affect the functionalities of the electronic devices. For example, the chalcogen vacancy defect[9] and its related complex[10] are able to activate additional defect-associated photoluminescence (PL) channels, which can decrease the intrinsic exciton PL efficiency. Charged impurity scattering due to either the ionized impurities inside TMDs[11] or charge traps at the surface of the substrate[12] has been proposed to be the dominant factor for the observed low room-temperature mobility in TMDs devices. Charge traps at the TMD-gate insulator interface have also been suggested as the cause of hysteresis in TMD-based field effect transistors[13]. Some defects can also serve as recombination centers, assisting the recombination of the photo-excited excitons/carriers in exfoliated few-layer and multilayer TMDs via Auger scattering[14-15].

Even though the defect trapping effect from oxygen impurities was proposed to be responsible for the previously observed phenomena in CVD-grown monolayer and multilayer $MoSe_2$[16], direct evidence and mechanisms about how these impurities are introduced into the samples, and more importantly, their effects on the electronic band structure and carrier dynamics are still not fully understood. In this letter, we intentionally created oxygen defects in exfoliated multilayer $MoSe_2$ by plasma irradiation, and then studied and verified the effect of the generated oxygen defects on the photo-excited carrier dynamics with femtosecond laser pump-probe spectroscopy. As expected, the optical signature of carriers being trapped by mid-gap defects is observed after plasma irradiation, and the defect-capturing-carrier effect increases with larger oxygen defect density. Our experimental findings, along with a band structure calculation with density functional theory, confirms the carrier-trapping role of oxygen impurities in TMD materials, and provides important implication for defect engineering in TMD-based devices.

Multilayer MoSe$_2$ samples were mechanically exfoliated from commercially purchased MoSe$_2$ crystal (2D Semiconductors) onto SiO$_2$/Si substrate using scotch tape. Since it has been found that the defect trapping exciton/carrier phenomenon occurs regardless of the sample thickness[16], indicating that the sample thickness is not a crucial factor, a relatively thick flake (80 nm thickness confirmed by AFM, see supplementary material) with lateral dimensions > 20 um was chosen in order to have enough sample area for the laser spot (20 um in diameter).

Plasma irradiation is an effective means to generate defects in TMD materials[17] or reduce the thickness of layered materials[18]. It has been demonstrated that oxygen impurity can be created in MoS$_2$ samples after Ar$^+$ plasma treatment[19]. Initially the Ar$^+$ plasma generates vacancies and new edges defects, then the vacancies and the new edges will be occupied or oxidized by oxygen atoms once exposed to air. Following this method, we performed two sequential Ar$^+$ plasma irradiations on our MoSe$_2$ sample using a reactive ion etcher (Plasma-Therm 790 RIE). The recipe used in the plasma treatment was 50 W power, 200 mTorr pressure, 20 sccm Ar flow rate, but with 30 s irradiation time for the 1$^{st}$ treatment and 300 s for the 2$^{nd}$ treatment. In order to verify the generation of defects, as well as identify the defect species, we measured the samples with X-ray photoelectron spectroscopy (XPS) after each plasma treatment, as shown in Figure 1. Before plasma irradiation, Mo3d$_{5/2}$ and Mo3d$_{3/2}$ peaks, located at 228.5 eV and 231.6 eV respectively, and Se3d$_{5/2}$ and Se3d$_{3/2}$ peaks, located at 54 eV and 54.8 eV respectively, can be observed. Theses peak positions agree with the reference values of Mo-Se bonding [20-21]. After the 1st plasma treatment (30s), four new peaks appear, located at around 232.3 eV, 235.4 eV, 58.6 eV and 59.4 eV respectively. Comparing with the literature results[22-23], two of these peaks originate from electrons in the Mo3d shell with j=5/2 and j=3/2 in Mo-O bonds, and the other two peaks from electrons in the Se3d shell with j=5/2 and j=3/2 in Se-O bonds, respectively. The observation of these peaks indicates that we have successfully created oxygen impurities with chemical bonds to both Mo and Se atoms in the material. After the 2$^{nd}$ plasma treatment, for which the irradiation time is 10 times than the 1$^{st}$ treatment, the relative intensities of the Mo-Se bonding peaks decreases while those of the Mo-O and Se-O bonding peaks increases. This indicates that more Mo-Se bonds were destroyed and more oxide bonds were created, i.e. the density of oxygen impurities increases, by the longer irradiation time.

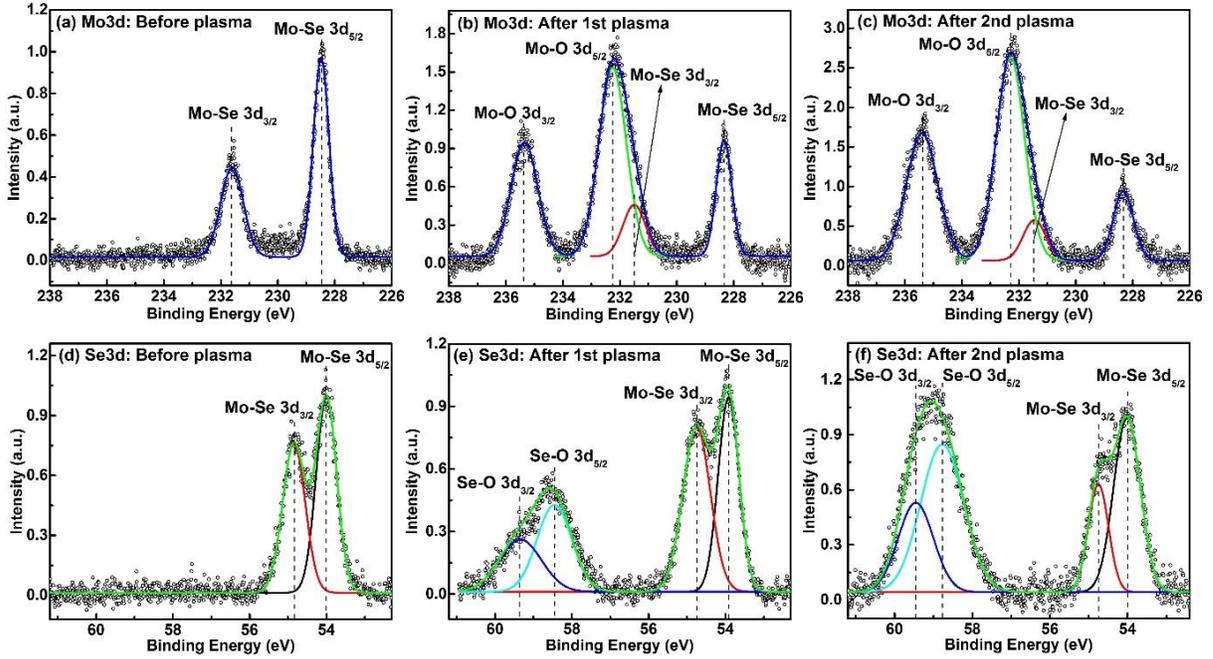

Figure 1. XPS data of MoSe$_2$ sample before and after plasma irradiation. Peaks of Mo-O and Se-O bonds can be clearly seen after plasma irradiation, and the relative intensity of these oxygen bonding peaks increase after the second plasma treatment.

The controllable generation of oxygen impurities allows us to investigate their effect on carrier dynamics by directly comparing the differential reflection signals before and after plasma treatments. We measured the differential reflection, $\Delta R/R_0=(R-R_0)/R_0$, with femtosecond laser pump-probe spectroscopy, where $R_0$ and $R$ are the reflections before and after the excitation by the pump pulse, respectively. The laser pulses are generated from a Ti: Sapphire femtosecond laser oscillator, with about 100 fs pulse width, 800 nm central wavelength, and 80 MHz repetition rate. Figure 2a shows the obtained $\Delta R/R_0$ signals at different pump fluences before plasma treatments. All the signals show negative exponential decays. Figure 2b illustrates the carrier generation and relaxation dynamics in bulk MoSe$_2$[24]. Since the energy difference between the conduction band minimum (CBM) and the valence band maximum (VBM) at the K point is around 1.5eV, we can generate and detect electrons and holes in the K valleys nearly-resonantly with 800nm pump and probe pulses, respectively. When probing resonantly, the reflection change is mainly due to the change in absorption caused by phase-space filling effect[25]. The reflection of multilayer $MoSe_2/SiO_2/Si$ structures can be calculated by the transfer matrix method and the derived expression is[26]:

$$R = \left|\frac{r_1 e^{i(\varphi_1+\varphi_2)}+r_2 e^{-i(\varphi_1-\varphi_2)}+r_3 e^{-i(\varphi_1+\varphi_2)}+r_1 r_2 r_3 e^{-i(\varphi_1-\varphi_2)}}{e^{i(\varphi_1+\varphi_2)}+r_1 r_2 e^{-i(\varphi_1-\varphi_2)}+r_1 r_3 e^{-i(\varphi_1+\varphi_2)}+r_2 r_3 e^{-i(\varphi_1-\varphi_2)}}\right|^2, \qquad (1)$$

where $r_1 = \frac{\tilde{n}_0-\tilde{n}_1}{\tilde{n}_0+\tilde{n}_1}$, $r_2 = \frac{\tilde{n}_1-\tilde{n}_2}{\tilde{n}_1+\tilde{n}_2}$, $r_3 = \frac{\tilde{n}_2-\tilde{n}_3}{\tilde{n}_2+\tilde{n}_3}$ are the complex amplitude of reflection coefficients for air/MoSe$_2$, MoSe$_2$/SiO$_2$, and SiO$_2$/Si interfaces, respectively; $\tilde{n}_i = n_i - i\kappa_i$ is the complex refractive index of each material (note that positive $\kappa$ stands for absorption), and $\varphi_i = 2\pi\tilde{n}_i d_i/\lambda$ is the complex phase shift due to a change in the optical path and the absorption in MoSe$_2$ or Si. Using Eq. (1), the relative reflection change, $\Delta R/R_0$, is plotted as a function of $\Delta\kappa/\kappa$ in the inset of Figure 2a. It can be seen that the reflection change is proportional to the absorption change with a positive slope. Therefore, the observed negative reflection change means that the absorption in the sample decreases after the pump excitation, due to the phase-space filling effect. When the excited carriers occupy the K valleys and reduce the relevant transitions due to Pauli blocking, absorption of the probe becomes less compared with that before pump excitation. In Figure 2a, there are two decay components in the negative reflection change signals, with the fast one only lasting for several ps and the slow one persisting for several hundred ps. The fast decay component can be attributed to carrier redistribution (also called thermalization and cooling) via intra-valley and inter-valley scatterings, which are accomplished by numerous processes of carrier-carrier scattering and carrier-phonon scattering, as shown in Figure 2b. The slow decay component reflects the direct recombination process of the carriers in the K valleys after the carrier redistribution, as also shown in Figure 2b[25].

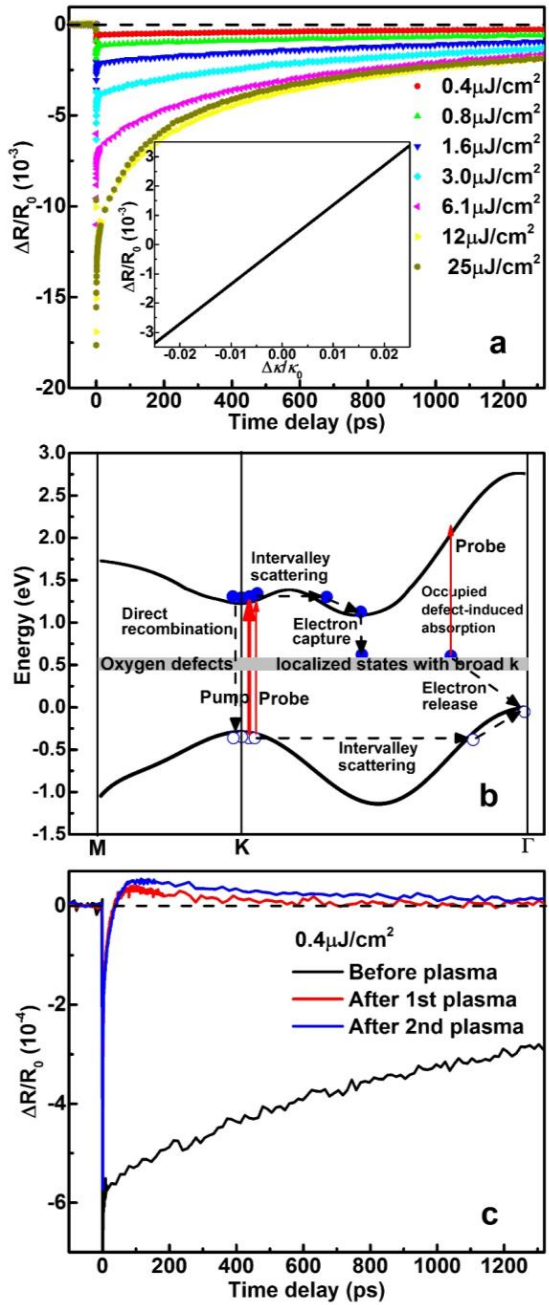
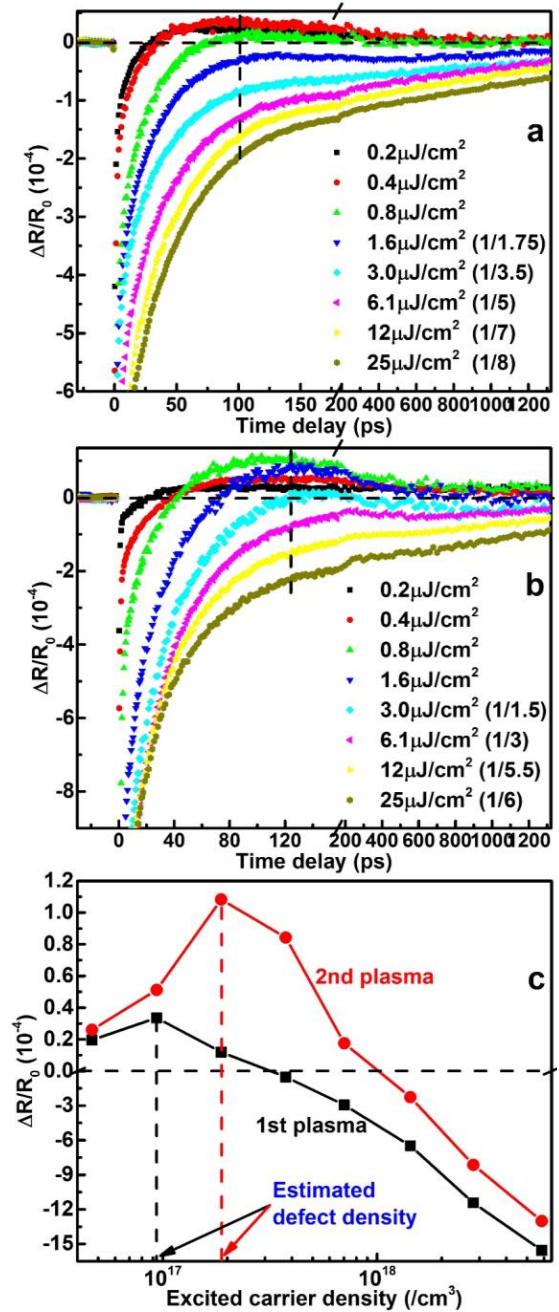

Figure 2. (a) Transient differential reflection signals $\Delta R/R_0$ measured at different pump fluences. (b) Band structure of multilayer $MoSe_2$ and schematic diagram of the photo-excited carrier dynamics, including carrier generation, inter-valley scattering, and carrier trapping by the mid-gap defects. Only electron capture is shown for simplicity. (c) Direct comparison of $\Delta R/R_0$ before and after plasma treatment at $0.4\mu J/cm^2$ pump fluence.

Figure 3. Transient differential reflection signals $\Delta R/R_0$ measured at different pump fluences (a) after the 1st and (b) the 2nd plasma irradiation. The signals at high fluences have been scaled by factors for a better view (indicated in the legend). (c) $\Delta R/R_0$ values at time delays marked with dashed lines in (a) and (b), as a function of excited carrier density.

The direct comparison of ΔR/R$_0$ signals before, and after the 1$^{st}$ and 2$^{nd}$ plasma treatments are shown in Figure 2c. At low pump fluences, one prominent feature of ΔR/R$_0$ after plasma irradiation is the sign change of the signals from negative to positive, indicating the absorption first decreases and then increases. Such sign-changing transient signal with the indication of an initial decrease followed by an enhancement in absorption has been observed in a number of defected materials, such as low temperature grown (LTG) traditional semiconductors by molecular beam epitaxy (LTG-GaAs[27-28], LTG-InP[29], LTG-InGaP[29]), and CVD grown layered MoSe$_2$ films[16], and well accepted as the signature of defect-capturing-carrier phenomenon. As shown in Figure 2b, when electrons are just excited into the conduction band, phase-space filling effect can cause absorption decrease. However, if the material has a considerable amount of mid-gap defects, the excited carriers will be quickly captured and trapped by those mid-gap defects. Once the electrons are captured, an additional absorption channel for the probe, i.e. the transition from the localized defect states which possess broad momenta to high energy states in the conduction band, becomes available, leading to an increase in the absorption. In our previous study[16], oxygen impurities have been proposed to be responsible for the carrier-trapping events, because the defect trapping effect could only be observed in CVD grown films but not in exfoliated flakes, and the oxygen content is only found in CVD as-grown films. Here, the fact that the defect-capturing-carrier signature can again be observed after the generation of oxygen impurities in the exfoliated sample by plasma irradiation, serves as strong evidence that oxygen impurities in MoSe$_2$ samples are indeed the effective carrier-trapping defects. While the negatively decaying black curve in Figure 2c shows the recombination of free K-valley carriers in the undamaged sample, the positively decaying red and blue signals in Figure 2c reflects the process of carrier releasing from the defect states back to the valence band, as shown in Figure 2b.

The ΔR/R$_0$ signals after the 1$^{st}$ and 2$^{nd}$ plasma treatments at different pump fluences are shown in Figure 3a and 3b, respectively. The competition between the trapped carriers and the excess free carriers, the defect saturation effect and the defect density can be investigated from the pump-fluences dependent ΔR/R$_0$ signals. Figure 3a and 3b show that the sign-changing signals can only be observed at low pump fluences. As pump power increases, the ΔR/R$_0$ signals will again become totally negative. This is because when the pump fluence is large enough, the excited carrier density will surpass the defect density, so all the defects will be occupied and excess free carriers will exist and take effect. In this case, the final $\Delta R/R_0$ signal is a result of two competing mechanisms: the trapped carriers that contribute positively to the signal via providing additional absorption channels; and the free carriers that contribute negatively to the signal through Pauli blocking effect. Obviously when the excited carrier density is much larger than the defect density, the negative component will be dominant and the signal will become totally negative, as shown by the signals taken at high fluences in Figure 3a and 3b.

Another interesting feature of the $\Delta R/R_0$ signals is the changing trend of the positive plateau values. As the pump fluence increases, the plateau value first increases then decreases, and the value around that time-delay finally decreases down to the negative region, as marked by the dashed lines in Figure 3a and Figure 3b. Using the transfer matrix method, the reflectance at the MoSe$_2$ surface ($R_0$), the transmittance into Si substrate (T), and hence the absorbance within the MoSe$_2$ (A=1-$R_0$-T) can be calculated[30]. With the obtained absorbance, we convert the pump fluence to the excited carrier density, and plot the $\Delta R/R_0$ values along the vertical dashed lines as a function of excited carrier density, as shown in Figure 3c. The initially increasing then decreasing trend can be seen very clearly. This trend actually reflects the defect saturation effect: Before defect saturation, more excited carriers will result in more occupied defects, and hence a larger positive signal; After defect saturation, more excited carrier will result in more excess free carriers which can compete with the trapped carriers and drag the total signal downward. As a result, more excited carrier will lead to more negative signal once the defects are saturated. Thus, the trend-turning point indicated by the dashed lines in Figure 3c actually correspond to the density of the available trapping-defects in the sample. The estimated defect density after the 1$^{st}$ and 2$^{nd}$ plasma treatments are $9.4 \times 10^{16}/cm^3$ and $1.8 \times 10^{17}/cm^3$, respectively. These results are consistent with the observation from the XPS data that after the 2$^{nd}$ plasma treatment, the oxygen impurities density increases. In other words, when the oxygen impurity density increases, the trapping defect density estimated from the $\Delta R/R_0$ signal also increase accordingly, which serves as another evidence for the carrier-trapping role of the oxygen impurities.

Finally, we want to discuss about the specific types of the oxygen impurities in MoSe$_2$. Although we have demonstrated that oxygen impurities are responsible for carrier trapping, it is still not clear which oxygen bonding is relevant, Mo-O (oxygen atoms occupying Se vacancies) or Se-O (oxygen atoms occupying Mo vacancies) or both? Shallow defects usually serve as donors or acceptors in a material, while deep mid-gap defects can serve as carrier trappers or recombination centers[31]. Thus, the key to answer this question is the defect energy position in the band structure, i.e., whether the bonding introduces mid-gap state in the band gap. Theoretical calculations with density functional theory (DFT) have shown that both molybdenum and chalcogen vacancies can induce mid-gap state within the bandgap[32-36]. However, the occupation of the chalcogen vacancy by oxygen atom (Mo-O bonding) can remove the mid-gap state and restore the band structure to a clean defect-free band gap[32-34]. In oxygen treated $MoS_2$ samples, substitution of chalcogen vacancy defects with oxygen atoms (Mo-O bonding) has been shown experimentally to improve the PL efficiency[32] and enhance carrier mobility[37]. Therefore, the question that remains is whether introducing oxygen atoms into the Mo vacancy sites to form Se-O bonding will also passivate the mid-gap states or give rise to more of them. We have performed DFT calculations to obtain the band structure of bulk $MoSe_2$ with O atoms occupying Mo vacancy sites. Details of DFT calculations can be found in the supplementary material. As can be seen in Figure 4, after O atoms occupy the Mo vacancy sites, additional defect states that located deep inside the original band gap of bulk $MoSe_2$ appear. These mid-gap states can typically serve as effective carrier trappers (note that recombination center is just a special case of carrier trapper with comparable electron and hole capturing rates)[31]. Therefore, considering together the DFT result and our experimental observations, we can conclude that oxygen impurities occupying Mo vacancy sites (and forming Se-O bonding) generate mid-gap defect states and play the role of carrier-trapper in $MoSe_2$ material. Our results imply that even the same defect species, e.g. oxygen atoms in this case, can have remarkably different effects on band structure and carrier dynamics in TMDs, depending on their position in the host lattice. When utilizing defect engineering to tune the optical and electrical properties of TMD materials, we need to know not only the defect species, but also how these defects interact with the host lattice, as both of these aspects are important to realize the desired functionality.

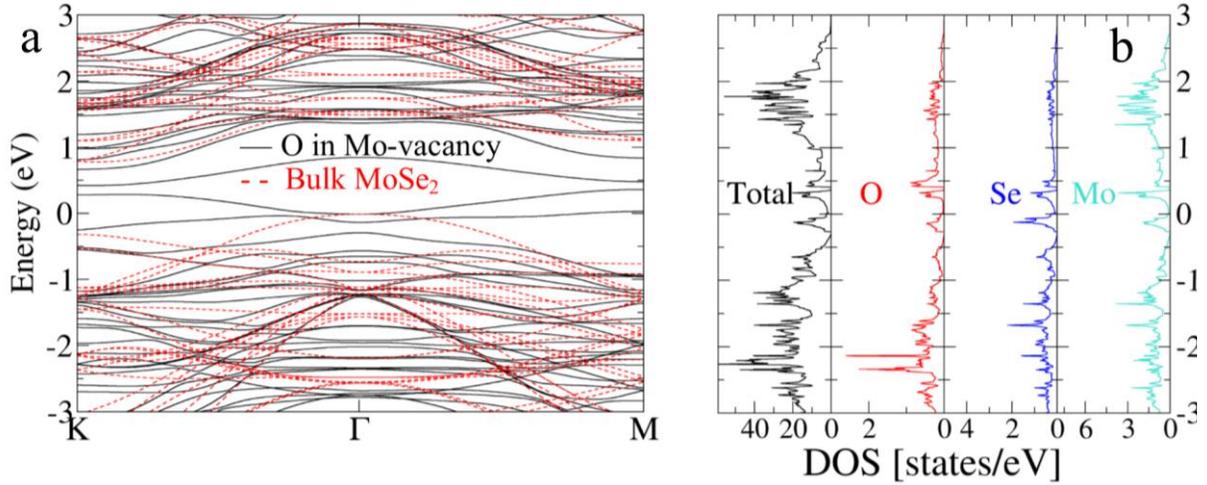

Figure 4. (a) DFT calculation of band structure of bulk $MoSe_2$ without defects and with oxygen atoms sitting in Mo sites. Mid-gap states can be seen clearly for the O in Mo-vacancy case. (b) Atom-projected density of state for $MoSe_2$ with oxygen atoms in Mo vacancy.

In summary, we have successfully created oxygen impurities in exfoliated multilayer $MoSe_2$ by $Ar^+$ plasma irradiations. The generation and increase of oxygen defects were confirmed by XPS. Transient reflection signals $\Delta R/R_0$ before plasma are composed of two negative fast and slow decays that reflect the inter-valley scattering and recombination of the excited carrier. The $\Delta R/R_0$ signals after plasma treatment exhibit a rapid sign change from negative to positive, a signature of defect trapping carriers, revealing the process of the excited carriers being captured by the created oxygen impurities. The density of oxygen impurities can be estimated from the $\Delta R/R_0$ values of the positive plateau, as a function of the density of excited carrier, and the result of increased defect density in the longer-irradiated sample agrees with the XPS observation. Moreover, DFT calculations show that oxygen atoms in Mo vacancy sites are responsible for the mid-gap defect states that accounts for the carrier trapping. Our findings shed light on the understanding of charge carrier-oxygen defect interactions in TMD materials, and provide important implications for defect-engineering in TMD-based devices.

## ASSOCIATED CONTENT

**Supporting Information**

The Supporting information is available on the ACS publications website.

The determination of MoSe2 flake thickness with AFM; and the details of DFT calculation for the band structure of MoSe2 with O atoms occupying Mo vacancy sites.

## AUTHOR INFORMATION
**Corresponding Author**

*E-mail: yaguo.wang@austin.utexas.edu

**Funding Sources**

National Science Foundation (NASCENT, Grant No. EEC-1160494; CAREER, Grant No. CBET-1351881); Department of Energy (SBIR/STTR, Grant No. DE-SC0013178); and DOD_ Army (Grant No. W911NF-16-1-0559).

**Notes**

The authors declare no competing financial interest.

## ACKNOWLEDGEMENTS

The authors acknowledge supports from National Science Foundation (NASCENT, Grant No. EEC-1160494; CAREER, Grant No. CBET-1351881), Department of Energy (SBIR/STTR, Grant No. DE-SC0013178), and DOD _ Army (Grant No. W911NF-16-1-0559). The authors thank Dr. Tiger Hu Tao and Shaoqing Zhang for the AFM measurements.

References

(1) Mak, K. F.; Lee, C.; Hone, J.; Shan, J.; Heinz, T. F. Atomically Thin $MoS_2$: A New Direct-Gap Semiconductor. *Phys. Rev. Lett.* **2010,** *105* (13), 136805.

(2) Mak, K. F.; He, K.; Lee, C.; Lee, G. H.; Hone, J.; Heinz, T. F.; Shan, J. Tightly Bound Trions in Monolayer $MoS_2$. *Nat. Mater.* **2013,** *12* (3), 207-211.

(3) Liu, H.; Neal, A. T.; Ye, P. D. Channel Length Scaling of $MoS_2$ MOSFETs. *ACS Nano* **2012,** *6* (10), 8563-8569.

(4) Xu, X.; Yao, W.; Xiao, D.; Heinz, T. F. Spin and Pseudospins in Layered Transition Metal Dichalcogenides. *Nature Physics* **2014,** *10* (5), 343-350.

(5) Ceballos, F.; Zhao, H. Ultrafast Laser Spectroscopy of Two-Dimensional Materials Beyond Graphene. *Advanced Functional Materials* **2017,** *27* (19).

(6) Wang, R.; Ruzicka, B. A.; Kumar, N.; Bellus, M. Z.; Chiu, H.-Y.; Zhao, H. Ultrafast and Spatially Resolved Studies of Charge Carriers in Atomically Thin Molybdenum Disulfide. *Phys. Rev. B* **2012,** *86* (4), 045406.

(7) Chen, K.; Yogeesh, M. N.; Huang, Y.; Zhang, S.; He, F.; Meng, X.; Fang, S.; Sheehan, N.; Tao, T. H.; Bank, S. R.; Lin, J.-F.; Akinwande, D.; Sutter, P.; Lai, T.; Wang, Y. Non-Destructive Measurement of Photoexcited Carrier Transport in Graphene with Ultrafast Grating Imaging Technique. *Carbon* **2016,** *107*, 233-239

(8) Strait, J. H.; Nene, P.; Rana, F. High Intrinsic Mobility and Ultrafast Carrier Dynamics in Multilayer Metal-Dichalcogenide $MoS_2$. *Phys. Rev. B* **2014,** *90* (24), 245402.

(9) Tongay, S.; Suh, J.; Ataca, C.; Fan, W.; Luce, A.; Kang, J. S.; Liu, J.; Ko, C.; Raghunathanan, R.; Zhou, J. Defects Activated Photoluminescence in Two-Dimensional Semiconductors: Interplay between Bound, Charged, and Free Excitons. *Sci. Rep.* **2013,** *3*, 2657.

(10) Wu, Z.; Luo, Z.; Shen, Y.; Zhao, W.; Wang, W.; Nan, H.; Guo, X.; Sun, L.; Wang, X.; You, Y. Defects as A Factor Limiting Carrier Mobility in WSe$_2$: A Spectroscopic Investigation. *Nano Res.* **2016,** *9* (12), 3622-3631.

(11) Ghatak, S.; Pal, A. N.; Ghosh, A. Nature of Electronic States in Atomically Thin MoS$_2$ Field-Effect Transistors. *ACS Nano* **2011,** *5* (10), 7707-7712.

(12) Ma, N.; Jena, D. Charge Scattering and Mobility in Atomically Thin Semiconductors. *Physical Review X* **2014,** *4* (1), 011043.

(13) Illarionov, Y. Y.; Rzepa, G.; Waltl, M.; Knobloch, T.; Grill, A.; Furchi, M. M.; Mueller, T.; Grasser, T. The Role of Charge Trapping In MoS$_2$/SiO$_2$ and MoS$_2$/hBN Field-Effect Transistors. *2D Mater.* **2016,** *3* (3), 035004.

(14) Wang, H.; Zhang, C.; Rana, F. Ultrafast Dynamics of Defect-Assisted Electron–Hole Recombination in Monolayer MoS$_2$. *Nano Lett.* **2014,** *15* (1), 339-345.

(15) Wang, H.; Zhang, C.; Rana, F. Surface Recombination Limited Lifetimes of Photoexcited Carriers in Few-Layer Transition Metal Dichalcogenide MoS$_2$. *Nano Lett.* **2015,** *15* (12), 8204-8210.

(16) Chen, K.; Ghosh, R.; Meng, X.; Roy, A.; Kim, J.-S.; He, F.; Mason, S. C.; Xu, X.; Lin, J.-F.; Akinwande, D. Experimental Evidence of Exciton Capture by Mid-Gap Defects in CVD Grown Monolayer MoSe$_2$. *npj 2D Materials and Applications* **2017,** *1* (1), 15.

(17) Chow, P. K.; Jacobs-Gedrim, R. B.; Gao, J.; Lu, T.-M.; Yu, B.; Terrones, H.; Koratkar, N. Defect-Induced Photoluminescence in Monolayer Semiconducting Transition Metal Dichalcogenides. *ACS Nano* **2015,** *9* (2), 1520-1527.

(18) Liu, Y.; Nan, H.; Wu, X.; Pan, W.; Wang, W.; Bai, J.; Zhao, W.; Sun, L.; Wang, X.; Ni, Z. Layer-By-Layer Thinning Of MoS$_2$ by Plasma. *ACS Nano* **2013,** *7* (5), 4202-4209.

(19) Tao, L.; Duan, X.; Wang, C.; Duan, X.; Wang, S. Plasma-Engineered MoS$_2$ Thin-Film as An Efficient Electrocatalyst for Hydrogen Evolution Reaction. *Chemical Communications* **2015,** *51* (35), 7470-7473.

(20) Vishwanath, S.; Liu, X.; Rouvimov, S.; Mende, P. C.; Azcatl, A.; McDonnell, S.; Wallace, R. M.; Feenstra, R. M.; Furdyna, J. K.; Jena, D. Comprehensive Structural and Optical Characterization of MBE Grown MoSe$_2$ on Graphite, CaF$_2$ and Graphene. *2D Mater.* **2015,** *2* (2), 024007.

(21) Azcatl, A.; Santosh, K.; Peng, X.; Lu, N.; McDonnell, S.; Qin, X.; de Dios, F.; Addou, R.; Kim, J.; Kim, M. J. HfO$_2$ on UV–O$_3$ Exposed Transition Metal Dichalcogenides: Interfacial Reactions Study. *2D Mater.* **2015,** *2* (1), 014004.

(22) Yao, W.; Iwai, H.; Ye, J. Effects of Molybdenum Substitution on The Photocatalytic Behavior of BiVO$_4$. *Dalton Transactions* **2008,** (11), 1426-1430.

(23) Chen, H.-Y.; Su, H.-C.; Chen, C.-H.; Liu, K.-L.; Tsai, C.-M.; Yen, S.-J.; Yew, T.-R. Indium-Doped Molybdenum Oxide as A New P-Type Transparent Conductive Oxide. *Journal of Materials Chemistry* **2011,** *21* (15), 5745-5752.

(24) Kumar, S.; Schwingenschlogl, U. Thermoelectric Response of Bulk and Monolayer MoSe$_2$ and WSe$_2$. *Chemistry of Materials* **2015,** *27* (4), 1278-1284.

(25) Kumar, N.; He, J.; He, D.; Wang, Y.; Zhao, H. Charge Carrier Dynamics in Bulk MoS$_2$ Crystal Studied by Transient Absorption Microscopy. *J. Appl. Phys.* **2013,** *113* (13), 133702.

(26) Blake, P.; Hill, E.; Neto, A. C.; Novoselov, K.; Jiang, D.; Yang, R.; Booth, T.; Geim, A. Making Graphene Visible. *Appl. Phys. Lett.* **2007,** *91* (6), 063124.

(27) Siegner, U.; Fluck, R.; Zhang, G.; Keller, U. Ultrafast High-Intensity Nonlinear Absorption Dynamics in Low-Temperature Grown Gallium Arsenide. *Appl. Phys. Lett.* **1996,** *69* (17), 2566-2568.

(28) Gupta, S.; Frankel, M.; Valdmanis, J.; Whitaker, J. F.; Mourou, G. A.; Smith, F.; Calawa, A. Subpicosecond Carrier Lifetime in GaAs Grown by Molecular Beam Epitaxy at Low Temperatures. *Appl. Phys. Lett.* **1991,** *59* (25), 3276-3278.

(29) Kostoulas, Y.; Waxer, L.; Walmsley, I.; Wicks, G.; Fauchet, P. Femtosecond Carrier Dynamics in Low-Temperature-Grown Indium Phosphide. *Appl. Phys. Lett.* **1995,** *66* (14), 1821-1823.

(30) Chen, K.; Sheehan, N.; He, F.; Meng, X.; Mason, S. C.; Bank, S. R.; Wang, Y. Measurement of Ambipolar Diffusion Coefficient of Photoexcited Carriers with Ultrafast Reflective Grating-Imaging Technique. *ACS Photonics* **2017,** *4(6)*, 1440-1446.

(31) Sze, S. M.; Ng, K. K. *Physics of semiconductor devices*, 3rd ed.; John wiley & sons: 2006, pp28.

(32) Su, W.; Jin, L.; Qu, X.; Huo, D.; Yang, L. Defect Passivation Induced Strong Photoluminescence Enhancement of Rhombic Monolayer $MoS_2$. *Phys. Chem. Chem. Phys.* **2016,** *18* (20), 14001-14006.

(33) Akdim, B.; Pachter, R.; Mou, S. Theoretical Analysis of the Combined Effects of Sulfur Vacancies and Analyte Adsorption on the Electronic Properties of Single-Layer $MoS_2$. *Nanotechnology* **2016,** *27* (18), 185701.

(34) Krivosheeva, A. V.; Shaposhnikov, V. L.; Borisenko, V. E.; Lazzari, J.-L.; Waileong, C.; Gusakova, J.; Tay, B. K. Theoretical Study of Defect Impact on Two-Dimensional $MoS_2$. *Journal of Semiconductors* **2015,** *36* (12), 122002.

(35) Valsaraj, A.; Chang, J.; Rai, A.; Register, L. F.; Banerjee, S. K. Theoretical and Experimental Investigation of Vacancy-Based Doping of Monolayer $MoS_2$ on Oxide. *2D Mater.* **2015,** *2* (4), 045009.

(36) Chiu, M.-H.; Li, M.-Y.; Zhang, W.; Hsu, W.-T.; Chang, W.-H.; Terrones, M.; Terrones, H.; Li, L.-J. Spectroscopic Signatures for Interlayer Coupling in $MoS_2$–$WSe_2$ Van der Waals Stacking. *ACS Nano* **2014,** *8* (9), 9649-9656.

(37) Nan, H.; Wu, Z.; Jiang, J.; Zafar, A.; You, Y.; Ni, Z. Improving the Electrical Performance of $MoS_2$ by Mild Oxygen Plasma Treatment. *Journal of Physics D: Applied Physics* **2017,** *50* (15), 154001.

# Supporting Information

# Carrier Trapping by Oxygen Impurities in Molybdenum Diselenide

*Ke Chen[1], Anupam Roy[2], Amritesh Rai[2], Amithraj Valsaraj[2], Xianghai Meng[1], Feng He[1,4], Xiaochuan Xu[3], Leonard F. Register[2], Sanjay Banerjee[2], Yaguo Wang[1,4]\**

1. Department of Mechanical Engineering, The University of Texas at Austin, Austin, TX 78712, USA
2. Microelectronics Research Center and Department of Electrical and Computer Engineering, The University of Texas at Austin, Austin, TX 78758, USA
3. Omega Optics, Inc., Austin, TX 78757, USA
4. Texas Materials Institute, The University of Texas at Austin, Austin, TX 78712, USA

*Corresponding Author: yaguo.wang@austin.utexas.edu

1. Determination of flake thickness with Atomic force microscope (AFM)

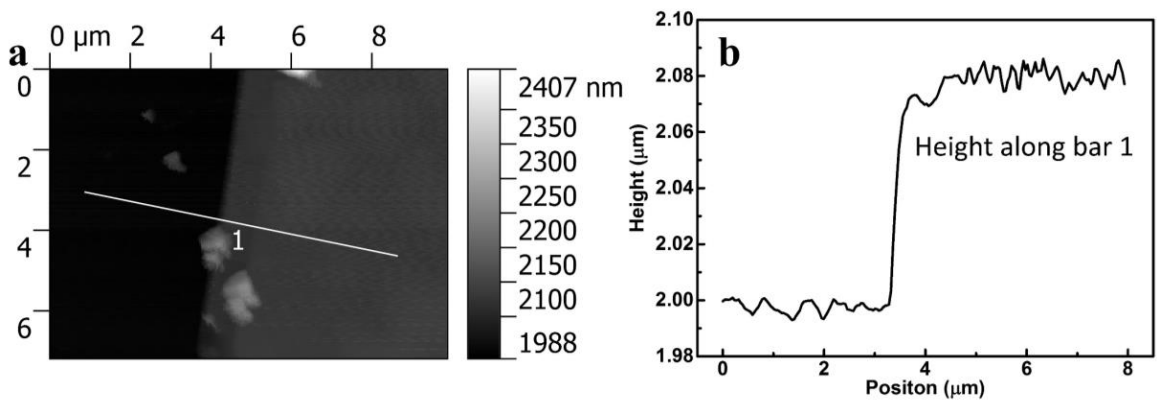

Figure s1. (a) Microscopic figure of the sample flake. (b) The height value along bar 1 across the sample boundary shown in (a), from which a thickness around 80nm is obtained.

2. Details of DFT band structure calculations

The DFT calculations were performed using the projector-augmented wave method with a plane-wave basis set as implemented in the Vienna *ab initio* simulation package (VASP)[1,2]. A kinetic energy cutoff of 400 eV was chosen. The k-mesh grid of 7x7x1 for the sampling of the first Brillouin zone of the supercell was selected according to Monkhorst-Pack type meshes, with the origin being at the Γ point for all computations except the band structure computation.[3] The local density approximation (LDA) was used for the exchange-correlation potential.[3,5] Van der Waals' forces were also simulated due to the absence of covalent bonding between the layers[4]. In our computations, we have adopted the DFT-D2 scheme to model the non-local dispersive forces wherein a semi-empirical correction is added to the conventional Kohn-Sham DFT theory[5]. A 2x2 supercell of bulk $MoSe_2$ was constructed with an O-atom incorporated into Mo-vacancy site. Atomistic relaxations were allowed to converge when the Hellmann-Feynman forces on the atoms were less than 0.005 eV/Ang. In previous studies, Mo-vacancy in TMDs has shown to introduce defect states in the nominal band gap of TMD material.[3,5] In our simulations, the highest occupied state of the system with vacancies serves as the zero- energy reference in these 0 K simulations.

References

1. Kresse, G.; Furthmüller, J. Efficient Iterative Schemes for Ab Initio Total-Energy Calculations Using A Plane-Wave Basis Set. *Phys. Rev. B* **1996**, 54, 11169.

2. Kresse, G.; Furthmüller, J. Efficiency of Ab-Initio Total Energy Calculations for Metals and Semiconductors Using A Plane-Wave Basis Set. *Comput. Mater. Sci.* **1996**, 6, 15.

3. Valsaraj, A.; Register, L. F.; Tutuc, E.; Banerjee, S. K. DFT Simulations of Inter-Graphene-Layer Coupling with Rotationally Misaligned hBN Tunnel Barriers in Graphene/hBN/Graphene Tunnel FETs. *J. Appl. Phys*. **2016**, 120, 134310.

4. McDonnell, S.; Brennan, B.; Azcatl, A.; Lu, N.; Dong, H.; Buie, C.; Kim, J.; Hinkle, C. L.; Kim, M. J.; Wallace, R. M. $HfO_2$ On $MoS_2$ by Atomic Layer Deposition: Adsorption Mechanisms and Thickness Scalability. *ACS Nano* **2013**, 7, 10354.

5. Grimme, S.; Antony, J.; Ehrlich, S.; Krieg, H. A Consistent and Accurate Ab Initio Parametrization of Density Functional Dispersion Correction (DFT-D) for the 94 Elements H-Pu. *J. Chem. Phys*. **2010**, 132, 154104.